\documentclass[prl,showpacs,showkeys,twocolumn]{revtex4}
\usepackage{graphicx}
\usepackage{amssymb}
\usepackage{amsmath}
\usepackage{subfigure}
\usepackage{bm}
\begin{document}
\title{Giant diamagnetism in half-metallic Co$_{2}$CrAl Heusler alloy}
\author{K. W. Kim$^{1}$} \author{J. Y. Rhee$^{2}$}\email{rheejy@skku.edu} \author{Y. V. Kudryavtsev$^{3}$, Y. H. Hyun$^{4}$, T. W. Eom$^{4}$}\author{Y. P. Lee$^{4}$}\email{yplee@hanyang.ac.kr}
\affiliation{$^{1}$Department of Physics, Sunmoon University, Asan 336-708, Korea\\
$^{2}$Department of Physics, Sungkyunkwan University, Suwon 440-746, Korea\\
$^{3}$Institute of Metal Physics, National Academy of Sciences of Ukraine, Kiev-142, Ukraine\\
$^{4}$Quantum Photonic Science Research Center and Department of
Physics, Hanyang University, Seoul, 133-791 Korea}
\date[]{Received  }

\begin{abstract}
A giant diamagnetism in the Co$_{2}$CrAl compounds, in both bulk
and thin film, below a certain temperature ($T_z$) was observed.
Above $T_z$, the compound behaves as an ordinary ferromagnet. The
diamagnetic alignment might be initiated by the Landau
diamagnetism because of the half-metallic properties and the
pinning of the diamagnetism is preserved by the peculiar
electronic structures.

\end{abstract}



\maketitle

Diamagnetism (DM) is a material property originating from the
response of an ensemble of charged particles governed by the
Faraday-Lenz law. Although it is ubiquitous to all materials, it
is usually masked by much stronger paramagnetism (PM) or
ferromagnetism (FM). Although some materials exhibit DM, the
strength is very weak. The only exceptions are superconductors in
which the magnetic field is completely expelled below the
superconducting transition temperature. This effect is known as
the Meissner effect.

According to the celebrated Bohr-van Leeuwen theorem \cite{ref1},
the orbital DM is impossible in classical thermodynamics, since
the 'skipping-orbit' of electrons near the surface cancels exactly
the bulk contribution to the diamagnetic moment. Later, by
applying quantum mechanics, Landau \cite{ref2} showed that the
bulk and the surface contributions were not balanced, for DM to
survive. Even for free-electron gas, however, the Landau DM is
only 1/3 of the Pauli PM. Consequently, the Landau DM cannot be
observed.

Although the Landau DM is too weak to be observed, a giant DM,
\emph{i.e.}, a magnetic susceptibility comparable to -1, is not
very rare. Claus and Veal \cite{ref3} observed a large negative
susceptibility in a weakly ferromagnetic Pd-0.5at.\% Fe alloy upon
field cooling (FC) under a very low field (a few tens of mG) and
concluded that the negative susceptibility was due to the
inhomogeneity of sample. This anomaly disappeared after removal of
the surface layer by grinding and chemical etching. The greatly
negative susceptibility was also noticed in a Pr-Al-Ni-Cu-Fe
metallic glass during the zero-field-cooled (ZFC) magnetization
measurement \cite{ref4}. It was due to the residual magnetic field
of magnet in the SQUID magnetometer, which was not supposed to
exist. A negative magnetic susceptibility observed in the FC
branch of spin- and charge-doped (Sr,La)(Ti,Co)O$_{3}$ was
explained by the existence of two magnetic sublattices and the
superposition of FM and PM to the magnetization \cite{ref5}. These
are not relevant to the Landau DM, but to some artifacts
originating from the measurements and the material preparation.

We found the peculiar magnetic properties of Co$_{2}$CrAl Heusler
alloy, both in bulk and thin films. The FC magnetization is
typical for the ferromagnetic material, while the ZFC
magnetization exhibits a giant negative magnetization, a hallmark
of extraordinary DM, at low temperatures and the magnetization
direction is flipped abruptly at a certain temperature upon
heating. The magnitude of negative magnetization is comparable to
that of the FC magnetization at the same temperature. The flipping
temperature strongly depends on the field strength. We attribute
this giant DM to the interplay between Landau DM and peculiar
electronic structures which are closely related to the
half-metallicity of Co$_{2}$CrAl Heusler alloy.

Bulk Co$_{2}$CrAl alloy was prepared by melting high purity
elements (99.99\%) in an arc furnace with a water-cooled Cu
hearth. The ingot was remelted 5 times and annealed at 1273 K for
10 h in vacuum to promote the volume homogeneity. No weight loss
after melting and heat treatment was observed. The alloy
composition was confirmed by x-ray fluorescence. Co$_{2}$CrAl
alloy films were prepared by flash evaporation of the crushed
alloy powders of 80 - 100 $\mu$m in diameter onto glass substrates
in a vacuum better than $2 \times 10^{-5}$ Pa. The alloy powders
were prepared from the same ingot of bulk Co$_{2}$CrAl alloy. To
enhance the crystallinity of film, the substrate was kept at 708 K
during deposition (film 1). Another film was deposited at 150 K
and post annealed at 760 K for 10 min (film 2). The thicknesses of
films are 135 and 153 nm for film 1 and 2, respectively. The
magnetic properties were investigated in a temperature range of $5
\leq T \leq 350$ K using a SQUID magnetometer for the samples
cooled in FC and ZFC modes, respectively.

\begin{figure}[tbp]
\scalebox{1}{\includegraphics{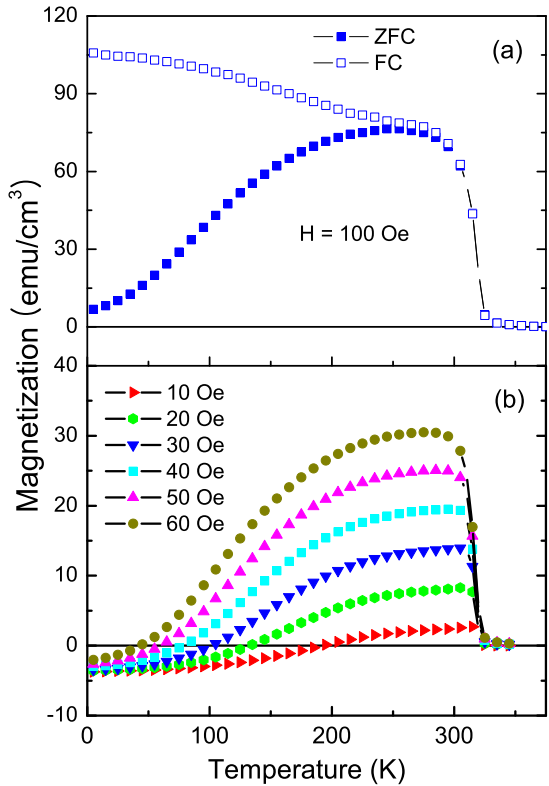}} \caption{(Color online)
Temperature dependence of the magnetization of the bulk
Co$_{2}$CrAl alloy. (a) FC and ZFC curves at \emph{H} = 100 Oe and
(b) ZFC curves at various applied fields.}\label{fig1}
\end{figure}

Figure \ref{fig1} presents the temperature dependence of
magnetization, $M(T)$, of the bulk sample. The Curie temperatures
$T_{\rm C}$ was 334 K. The $M(T)$ behavior of the bulk sample
measured at 100 Oe in the FC branch displays that of a typical
ferromagnet and the ZFC one is also typical for a ferromagnet with
some local structural disorder [Fig. \ref{fig1}(a)]. However, as
the external magnetic field decreases below 100 Oe, the ZFC curves
show significantly diamagnetic behavior at low temperatures. The
magnetization changes its sign at a certain temperature designated
as $T_z$. As the strength of measuring field increases, $T_z$
decreases. Since the investigated sample did not exhibit any
superconducting behavior down to 5 K, we can safely exclude the
superconductivity as the origin for the observed DM.

\begin{figure}[tbp]
\scalebox{1.0}{\includegraphics{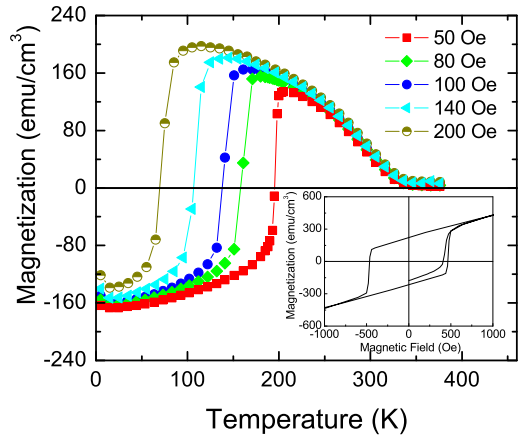}} \caption{(Color online)
Dependence of ZFC $M(T)$ for film 1 at various applied magnetic
fields. The inset shows the initial rising curve of film
1.}\label{fig2}
\end{figure}

The diamagnetic behavior in the ZFC mode at low temperatures is
more prominent for the thin films. In Fig. \ref{fig2} the ZFC
magnetization curves for film 1 at various magnetic fields are
displayed. In the thin-film samples, the strength of DM is nearly
comparable to that of the FC magnetization at the same
temperature.

In Ref. 3, a large negative susceptibility observed in the FC
branch at very low fields was due to the inhomogeneity of sample.
In our case the large DM was found in the ZFC branch and the
applied field for measurement was much larger than that of Ref. 3.
Therefore, the inhomogeneity of sample is excluded as the reason
for the observed diamagnetic behavior.

Since a non-zero but negative residual field of the
superconducting magnet during cooling is another possible reason
for the observed DM \cite{ref4}, we have checked the residual
field. It was +1.5 Oe, which is positive. Furthermore, if the
direction of applied field is flipped during the measurement after
ZFC, a large positive magnetization was observed and became
negative above $T_z$. Therefore, the observed giant DM is not due
to the residual field of SQUID magnetometer during cooling.

As aforementioned, every material can exhibit a diamagnetic
behavior. Once the magnetic field lines start to penetrate a
metallic sample, there are induced currents on the sample surface.
According to the Faraday-Lenz law, the induced current generates a
magnetic field opposing the applied field. If the induced currents
persist, they would expel the magnetic field completely out of the
interior of sample. In the normal metal, however, the finite
resistance kills the current instantly and, consequently, the
magnetic field can penetrate into the interior of sample.
Therefore, the DM is hardly observed in most of metallic
materials.

If a magnetic field is applied to a metal, the majority (minority)
bands move downward (upward) since the energy
-\textbf{$\mu$}$\cdot$\textbf{H} is added to the system, where
$\mu$ is the magnetic moment of an electron and \textbf{H} the
applied magnetic field. Accordingly, the Fermi level, more
precisely, the chemical potential for the majority (minority) band
moves downward (upward). Then, to equilibrate the chemical
potential, the electrons in the minority bands near the Fermi
level flip their spins and fill up the majority bands until the
Fermi level becomes uniform. This is the well-known Pauli PM,
which is three times stronger than the Landau DM even in
free-electron gas. In the ordinary ferromagnet, when a magnetic
field is applied to measure the magnetization, the Pauli
paramagnetic response would lead to the alignment of magnetic
moments inside the sample along the applied field.

On the other hand, in half metals, such as Co$_{2}$CrAl, the Pauli
PM cannot occur since there is no Fermi-level imbalance under an
applied field because of the absence of minority bands near the
Fermi level in Co$_{2}$CrAl. Therefore, the Landau DM might be
disclosed. This initial DM somehow pins the magnetic moments in
the interior of sample opposing the applied field, resulting in a
huge negative magnetization, as observed in our sample. The
apparent DM is destroyed by the applied field with a strength
higher than a certain value since the sample is supposed to be a
ferromagnet and, hence, the higher the applied field, the stronger
the tendency for the alignment of magnetic moments along the
applied field. Once the whole magnetic moments in the sample are
completely aligned, the Landau DM would be masked by the
ferromagnetic response and, consequently, the sample behaves as an
ordinary ferromagnet. This is clearly seen in the initial rising
curves of hysteresis of the bulk and the film samples (see the
inset of Fig. \ref{fig2}). The initial rising curves start with a
negative magnetization and become positive above a certain
critical magnetic field. After the positive magnetization is
achieved, the sample comes to be an ordinary ferromagnet.

The decrease of $T_z$ with increasing applied field can be
understood as follows. The magnetic moments of individual domains
are well aligned opposite to the applied field and are `frozen' at
low temperatures. Therefore, it is hard to flip the magnetization
to align along the applied field. The energy required to flip the
magnetization of a single domain is $\Delta E = K_A V$, where
$K_{A}$ is the magnetic anisotropy constant and \emph{V} is the
volume of domain. This energy can be supplied thermally. As
temperature increases, the magnetic moments tend to align along
the applied field more easily since the magnetic moments are more
agitated owing to the supplied extra thermal energy. This leads to
the decrease in the magnitude of negative magnetization. If
temperature reaches a certain temperature, the magnetic moments
flip their directions as a whole. When an external field is
applied, the energy barrier for magnetization flipping is reduced
to be $\Delta E = K_A V\left( {1 - \frac{{M_S H_C }}{{2K_A }}}
\right)$, where $M_{s}$ is the saturation magnetization and
$H_{c}$ is the coercive field. According to the N$\acute{{\rm
e}}$el-Brown model of magnetization reversal \cite{ref6}
\begin{equation}\label{eq1}
\frac{{H_c \left( T \right)}}{{H_c \left( 0 \right)}} \cong \left[
{1 - \left({\frac{T}{{T_0 }}} \right)^\alpha  } \right],
\end{equation}
where $H_{\rm{c}} \left( 0 \right) \equiv \frac{{M_{\rm{s}}
}}{{2K_{\rm{A}} }}$ is the coercive field at 0 K, $T_0  =
\frac{{K_{\rm{A}} V}}{{k_{\rm{B}} }}$ ($k_B$ is the Boltzmann
constant), and $\alpha$ a constant which depends on the average
angle between magnetic moments and applied field \cite{ref6-1}. At
$T = T_0$, the coercive field is zero.

$T_z$ is the temperature at which the magnetization reversal takes
place under a given magnetic field. Therefore, we can treat the
applied field as the coercive field at $T = T_z$. $\alpha$ is a
fitting parameter: $\alpha = 0.602$ for the bulk sample and 0.522
for film 2, while 0.254 for film 1. Although $\alpha$ is usually
close to 2/3, it can be as small as 0.5, depending on the average
angle between magnetic moments and the applied field. Therefore,
the magnitudes of $\alpha$ for the bulk and film 2 are reasonable
in a sense that the N$\acute{{\rm e}}$el-Brown model is applied,
however, it is too small for film 1. The x-ray diffraction
investigation (see Fig. 2 in Ref. 9) reveals that film 1 has a
significantly worse crystallinity than film 2. Film 1 has a
smaller grain size, and a great amount of intergranular amorphous
phase which plays an important role in determining the coercive
field \cite{ref10-1}, resulting in a different temperature
dependence of the coercivity from that of the bulk and film 2.

\begin{figure}[tbp]
\scalebox{.3}{\includegraphics{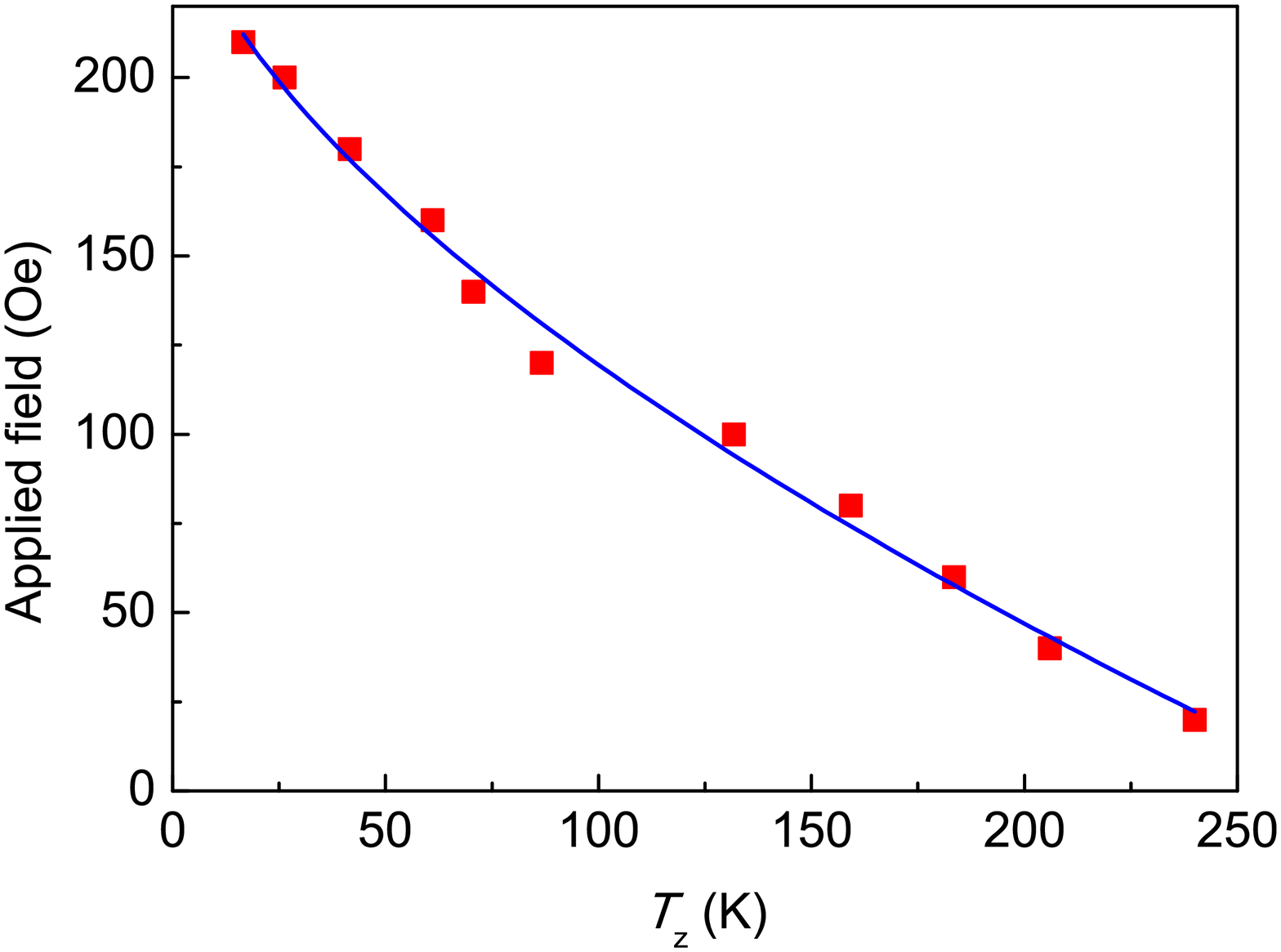}} \caption{(Color online)
Applied field as a function of $T_z$ for the bulk sample. Solid
lines are the fitting result.}\label{fig3}
\end{figure}

Even if we accept the scenario explained here, one might ask
another question, such as ``\emph{What kind of `force' can
initially hold or pin the magnetization opposite to the applied
field?}" We argue that the pinning of diamagnetic magnetization is
due to the peculiar electronic structures. We calculated the
electronic structures of Co$_{2}$CrAl compound under various
conditions by using the \emph{WIEN2k} package \cite{ref7}. For the
exchange-correlation functional, the
generalized-gradient-approximation version of Perdew \emph{et al.}
\cite{ref9} was used. The \emph{WIEN2k} package has a special
feature that the magnetic field can be added in the course of
self-consistent iterations and the direction of applied field can
be virtually arbitrary. Since the spin-orbit coupling is crucial
for this calculation, it is included. Since the magnitude of
energy difference we deal with is very tiny, the whole reciprocal
unit cell was divided into $50 \times 50 \times 50$
parallelepipeds to safely handle the subtle energy difference. We
also vary the direction of applied field.

\begin{figure}[tbp]
\scalebox{.35}{\includegraphics{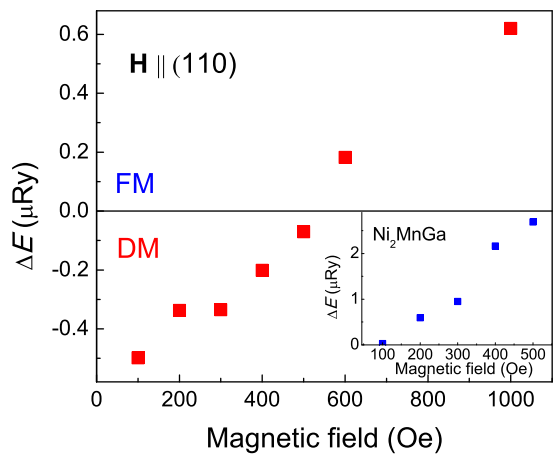}} \caption{(Color online)
Total energy difference between ferromagnetic and diamagnetic
states. Inset shows the same except for Ni$_{2}$MnGa.}\label{fig4}
\end{figure}

Once the magnetization direction is chosen, say, in the (001)
direction, the field is applied in the same (opposite) direction.
Hereafter, we will refer it to the FM (DM) magnetization. In the
ordinary ferromagnetic material, when an external magnetic field
is applied, the total energy is lower than the case without the
applied field by aligning the magnetic moments along the field. If
the magnetic field is applied opposite to the magnetization
direction, the opposite phenomenon happens and the total energy
becomes higher. As seen in the inset of Fig. \ref{fig4}, the total
energy difference of Ni$_{2}$MnGa compound is always positive and
increases according to the field strength. The total energy
difference $\Delta E$ is defined as $\Delta E \equiv
E_{{\rm{tot}}{\rm{.DM}}}  - E_{{\rm{tot}}{\rm{.FM}}}$. Positive
implies that the ferromagnetic state is more stable than the
diamagnetic one. Therefore, in the ordinary ferromagnet, the
diamagnetic alignment is impossible.

The situation becomes quite different if the same calculation is
done for the Co$_{2}$CrAl compound. In Fig. \ref{fig4}, the total
energy of the DM is lower than the FM when the applied field
strength is smaller than 500 Oe. Our calculational results also
reveal that the diamagnetic state is the ground state when an
applied field is lower than 500 Oe at 0 K. This peculiar
electronic structures of Co$_{2}$CrAl compound are the source of
pinning the DM unless a sufficient thermal energy is supplied to
agitate the magnetic moments to finally align along the applied
field. Even though the diamagnetic response might be small, it is
strong enough to pin the magnetic moments within the thin-film
sample, similar to the case exhibited in Ref. 3.

We observed a giant DM in the Co$_{2}$CrAl compound at low
temperatures. The phenomenon is persistent up to a certain
temperature $T_z$ which increases as the magnetic field strength
decreases. The diamagnetic alignment at low \emph{T} might be
initiated by the Landau DM, owing to the absence of minority bands
at the Fermi level, and the pinning of the DM is preserved by the
peculiar electronic structures of the compound. Although the
celebrated Landau DM is predicted more than 70 years ago, it had
not been possible to be observed since it is masked by stronger PM
or even FM. The interplay between half-metallicity and peculiar
electronic structures of the Co$_{2}$CrAl compound enables us to
observe this old puzzle of magnetism.

\acknowledgments This work was supported by MEST/KOSEF through the
Quantum Photonic Science Research Center, Korea, and by KOSEF
grant funded by the Korea government (MEST) (No.
R01-2007-000-20974-0).


\begin{thebibliography}{}
\bibitem{ref1} N. Bohr, Dissertation, Copenhagen (1911); J. N. Van Leeuwen, J. Phys. Radium \textbf{2}, 361-377 (1921).
\bibitem{ref2} L. D. Landau, Z. Phys. \textbf{64}, 629 (1930).
\bibitem{ref3}  H. Claus and B. W. Veal, Phys. Rev. B \textbf{56}, 872 (1997).
\bibitem{ref4} Y. T. Wang \emph{et al.}, Appl. Phys. Lett. \textbf{85}, 2881 (2004).
\bibitem{ref5} J. Yang \emph{et al.}, Appl. Phys. Lett. \textbf{91}, 052502 (2007).
\bibitem{ref6} W. Wernsdorfer \emph{et al.}, Phys. Rev. Lett. \textbf{78}, 1791 (1997).
\bibitem{ref6-1} M. P. Sharrock, IEEE Trans. Magn. \textbf{44}, 4414 (1999).
\bibitem{ref6-2} H. Pfeiffer, Phys. Stat. Sol. \textbf{118}, 295 (1990).
\bibitem{ref10} Y. V. Kudryavtsev \emph{et al.}, Phys. Rev. B \textbf{77}, 195104 (2008);
    K. W. Kim \emph{et al.}, J. Korean Phys. Soc. \textbf{53}, 2475
    (2008).
\bibitem{ref10-1} K. Suzuki and J.M. Cadogan, Phil. Mag. Lett. \textbf{77}, 371 (1998).
\bibitem{ref7} P. Blaha \emph{et al.}, \emph{WIEN2k, An Augmented Plane Wave + Local Orbitals Program for Calculating Crystal Properties} (Karlheinz Schwarz, Techn. Universit$\ddot{\rm a}$t Wien, Wien, Austria, 2001).
\bibitem{ref9} J. P. Perdew \emph{et al.}, Phys. Rev. Lett. \textbf{77}, 3865 (1996).

\end{thebibliography}
\end{document}